\documentclass[journal,twoside]{IEEEtran}
\makeatletter
\def\markboth#1#2{\def\leftmark{\@IEEEcompsoconly{\sffamily}\MakeUppercase{\protect#1}}%
\def\rightmark{\@IEEEcompsoconly{\sffamily}\MakeUppercase{\protect#2}}}
\makeatother

\usepackage[numbers,sort&compress]{natbib}

\usepackage{hyperref}
\usepackage{color}

\ifCLASSINFOpdf
  \usepackage[pdftex]{graphicx}
\else
\fi
\usepackage[cmex10]{amsmath}
\usepackage[caption=false,font=footnotesize]{subfig}
\usepackage[american]{babel}
\usepackage{fixltx2e}

\begin{document}
\title{A Comprehensive 3D Framework for Automatic Quantification of Late Gadolinium Enhanced Cardiac Magnetic Resonance Images}

\author{Dong~Wei$^*$,
        Ying~Sun,~\IEEEmembership{Member,~IEEE,}
        Sim-Heng~Ong,~\IEEEmembership{Member,~IEEE,}
        Ping~Chai,
        Lynette~L.~Teo,
        and~Adrian~F.~Low% <-this % stops a space
\thanks{D. Wei$^*$, Y. Sun and S. H. Ong are with the Department of Electrical and Computer Engineering, National University of Singapore, 4 Engineering Drive 3, Singapore 117576. email: \{dongwei, elesuny, eleongsh\}@nus.edu.sg}% <-this % stops a space
\thanks{P. Chai and A.F. Low are with the Cardiac Department, National University Heart Centre, 5 Lower Kent Ridge Road, Singapore 119074. email: \{ping\_chai, \mbox{adrian\_low}\}@nuhs.edu.sg}% <-this % stops a space
\thanks{L.L. Teo is with the Department of Diagnostic Imaging, National University Hospital, 5 Lower Kent Ridge Road, Singapore 119074. email: lynette\_ls\_teo@nuhs.edu.sg}
%\thanks{Manuscript received \ldots; revised \ldots}
}


\markboth{IEEE TRANSACTIONS ON BIOMEDICAL ENGINEERING, PREPRINT Jan 25, 2013}%
{Wei \MakeLowercase{\textit{et al.}}: A Comprehensive 3D Framework for Quantification of LGE Cardiac MR Images}



\maketitle

\begin{abstract}
%\boldmath
Late gadolinium enhanced (LGE) cardiac magnetic resonance (CMR) can directly visualize non-viable myocardium with hyper-enhanced intensities with respect to normal myocardium.
For heart-attack patients, it is crucial to facilitate the decision of appropriate therapy by analyzing and quantifying their LGE CMR images.
To achieve accurate quantification, LGE CMR images need to be processed in two steps: segmentation of the myocardium followed by classification of infarcts within the segmented myocardium.
However, automatic segmentation is difficult usually due to the intensity heterogeneity of the myocardium and intensity similarity between the infarcts and blood pool.
Besides, the slices of an LGE CMR dataset often suffer from spatial and intensity distortions, causing further difficulties in segmentation and classification.
In this paper we present a comprehensive 3D framework for automatic quantification of LGE CMR images.
In this framework, myocardium is segmented with a novel method that deforms coupled endocardial and epicardial meshes and combines information in both short- and long-axis slices, while infarcts are classified with a graph-cut algorithm incorporating intensity and spatial information.
Moreover, both spatial and intensity distortions are effectively corrected with specially designed countermeasures.
Experiments with 20 sets of real patient data show visually good segmentation and classification results that are quantitatively in strong agreement with those manually obtained by experts.
\end{abstract}


% \fi
\IEEEpeerreviewmaketitle

\section{Introduction}
%\IEEEPARstart{T}{his} demo file is intended to serve as a ``starter file''
%for IEEE journal papers produced under \LaTeX\ using
%IEEEtran.cls version 1.7 and later.

%\IEEEpubidadjcol

\IEEEPARstart{I}{schemic} heart disease is one of the leading causes of death in western countries \cite{kishore2011global}.
It refers to ischemia of the myocardium due to stenosis of the coronary arteries, which supply oxygenated blood to the myocardium.
If a stenosis develops to completely occlude the vessel, the patient undergoes a myocardial \emph{infarction}, i.e., heart attack.
The part of the myocardium which undergoes a prolonged shortage of oxygen is damaged and can be either non-viable or hibernating after the infarction.
Since the hibernating myocardium has the potential to resume contraction after re-vascularisation \cite{hibernatingMyocardium1989}, viability assessment of the myocardium is essential for diagnosis and therapy planning.

\begin{figure}
    \centerline{
    \hfil
    \subfloat[]{
        \includegraphics[width=.3\columnwidth]{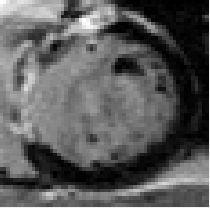}
        \label{fig_first_case}
    }
    \hfil
    \subfloat[]{
        \includegraphics[width=.3\columnwidth]{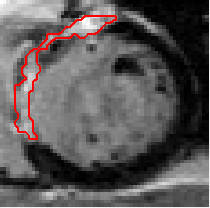}
        \label{fig_second_case}
    }
    \hfil
    \subfloat[]{
        \includegraphics[width=.3\columnwidth]{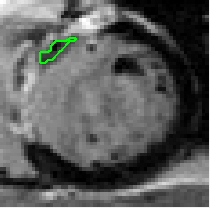}
        \label{fig_second_case}
    }
    \hfil
    }\caption{An example LGE image: (a) the original image; (b) the hyper-enhanced infarct is outlined. (c) the MVO is outlined.}\label{fig:LGEillustration4Introduction}
\end{figure}

Among various cardiac scans, late gadolinium enhanced (LGE) cardiac magnetic resonance (CMR) imaging can directly visualize and thus discriminate non-viable myocardium (known as infarcts) from normal myocardium \cite{LGEmedical1999}.
In an LGE CMR scan, a gadolinium-based contrast agent is injected and a single-frame sequence is acquired 10 to 20 minutes later, by which time infarcts exhibit hyper-enhanced intensities compared to healthy myocardium.
This phenomenon has been hypothesized to be the result of delayed wash-out kinetics of the contrast agent in non-viable myocardium.
Post-mortem histologic staining of the myocardium using animal models has shown that the hyper-enhanced regions in LGE CMR images correlate well with the location and extent of non-viable tissue \cite{LGEmedical1999,fieno2000contrast}.
However, there is one exception called \emph{microvascular obstruction} (MVO).
MVO refers to the no-reflow phenomenon in which the involved sub-endocardial infarcts do not get enhanced and appear as dark as normal myocardium.
It is mostly observed in acute infarctions~\cite{MVO2010}.
An LGE image with a hyper-enhanced infarct and MVO is shown in Fig.~\ref{fig:LGEillustration4Introduction}.
LGE CMR is usually triggered with electrocardiography (ECG) gating at mid-diastole of the cardiac cycle.

% An example of a double column floating figure using two subfigures.
% (The subfig.sty package must be loaded for this to work.)
% The subfigure \label commands are set within each subfloat command, the
% \label for the overall figure must come after \caption.
% \hfil must be used as a separator to get equal spacing.
% The subfigure.sty package works much the same way, except \subfigure is
% used instead of \subfloat.

%
% Note that often IEEE papers with subfigures do not employ subfigure
% captions (using the optional argument to \subfloat), but instead will
% reference/describe all of them (a), (b), etc., within the main caption.

Quantification of LGE CMR images can be divided into two stages: segmentation of the myocardium and classification of infarcts within the segmented myocardium.
Although the quantification can be done manually by experts, it is not only time-consuming but also subject to inter- and intra-observer variation.
Therefore, computer aided \mbox{(semi-)} automatic techniques are highly desired.
However, the automation is not straightforward.
First, automatic segmentation of the myocardium is often difficult due to the intensity heterogeneity of the myocardium and intensity similarity between the infarcts and blood pool (BP).
Second, misclassification of infarcts can happen because of the intensity inconsistencies and misalignment artifacts of a stack of slices, image noise and other artifacts.

Our contribution in this paper is a complete and comprehensive 3D framework for automatic quantification of LGE CMR images, which is rare in the literature.
This framework includes two crucial pre-processing measures -- corrections of misalignment artifacts and intensity inconsistencies -- that are omitted or improperly handled in previous works.
Furthermore, both segmentation and classification stages in the framework are 3D in nature, hence utilizing intensity and spatial continuity beyond individual 2D images.
The rest of this paper is organised as follows.
Section \ref{sec:background} introduces the background knowledge and reviews related works.
Section \ref{sec:method} describes the methods used in every step of our 3D framework.
The experimental results are presented and discussed in Section \ref{sec:results}, followed by the conclusions in Section \ref{sec:conclusion}.

\section{Background and Related Works}
\label{sec:background}

\subsection{Misalignment Correction}

\begin{figure}
  \centering
  \includegraphics[width=.95\columnwidth]{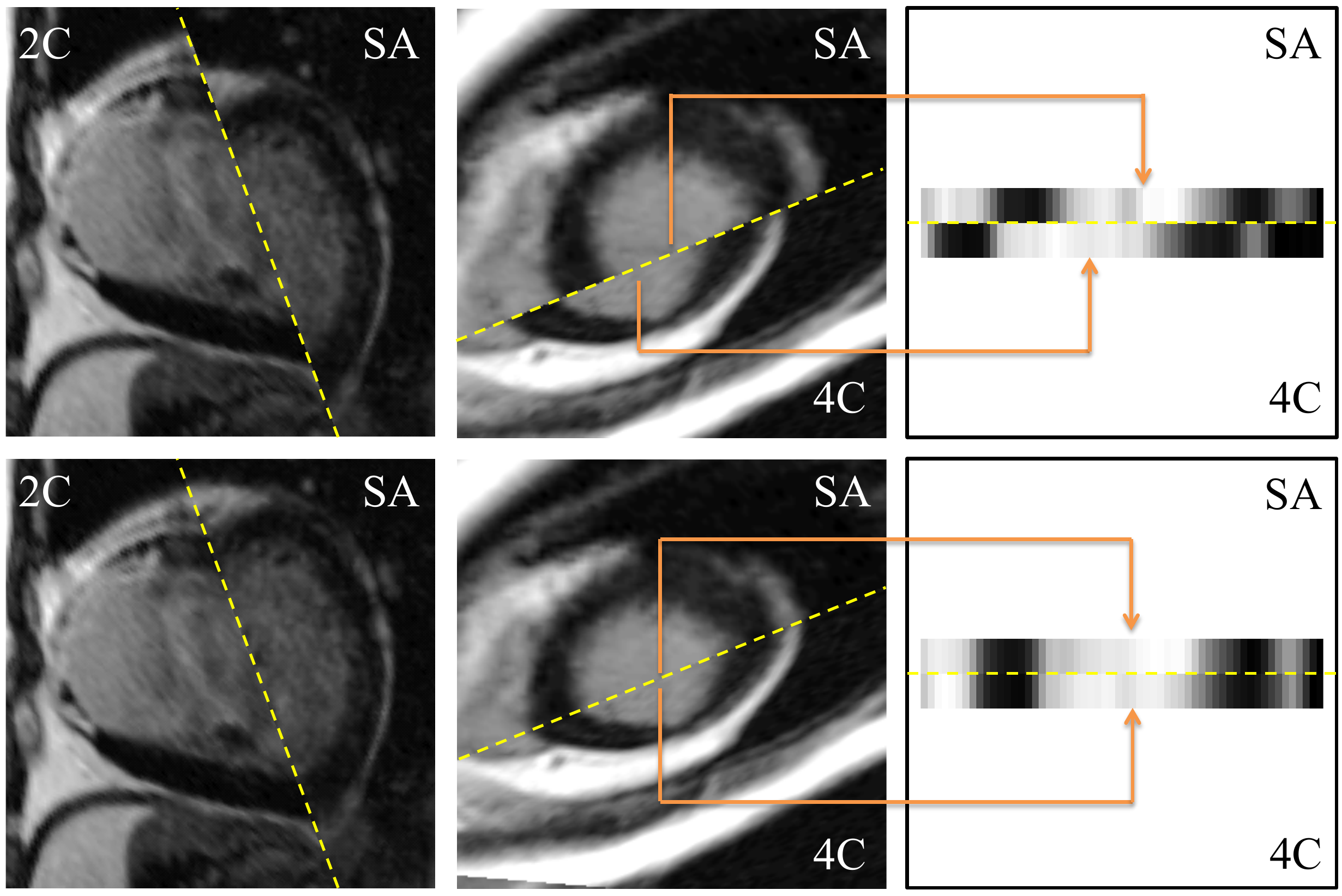}\\
  \caption{First row: misalignment artifacts illustrated with intersections between SA and LA slices.
  Second row: the same slices after misalignment correction.
  The right column shows sampled line segments along the intersection line from the slices in the mid column.
  4C: 4-chamber; 2C: 2-chamber.}\label{fig:misalignment_n_correction_n_intersection}
\end{figure}

Misalignment artifacts are common for multi-slice CMR data.
Clinically, a few consecutive short-axis (SA) slices and 1-3 standard long-axis (LA) slices are acquired.
Slices are usually acquired in multiple breath-holds, introducing differences in lung volume which in turn cause displacement of the heart \cite{heartMotionStudy}.
Misalignment artifacts can also be caused by patient motion and heartbeat;
the former can be considered similar to the breath-related artifacts, while the latter is often neglected due to the use of ECG gating.
Misalignment artifacts cause slices to move away from their true positions relative to each other (Fig.~\ref{fig:misalignment_n_correction_n_intersection}, first row) and, if not tackled, may lead to distortion of the 3D reconstruction from the multi-slice data.

% Reminder: the "draftcls" or "draftclsnofoot", not "draft", class
% option should be used if it is desired that the figures are to be
% displayed while in draft mode.

% where an .eps filename suffix will be assumed under latex,
% and a .pdf suffix will be assumed for pdflatex; or what has been declared
% via \DeclareGraphicsExtensions.
% Note that IEEE typically puts floats only at the top, even when this
% results in a large percentage of a column being occupied by floats.

Researchers have proposed various approaches to realign a set of CMR slices.
\citet{A_Comprehensive_Approach} registered LGE slices to a whole-heart coronary angiography volume slice by slice by maximizing similarity measures in the overlapping image parts.
However, the high resolution whole-heart volume data is not widely available under clinical settings.
\citet{alignManyLA2004} proposed a method more suitable for clinically acquired CMR, where slices were realigned by maximizing normalized mutual information (NMI) of the intersecting parts between slices.
However, no regularization term was included to avoid unrealistic movements.
Recently, \citet{alignEqualSpacing2010} introduced an effective regularization term favoring original spacing among SA slices, but its underlying assumption that the original spacing be correct is sometimes broken by superior-inferior motion of the heart.
A common deficiency of \cite{alignEqualSpacing2010,alignManyLA2004} is that they merely used similarity at the intersections between slices but ignored the anatomical continuity of the heart within the stack of SA slices.
Furthermore, both \cite{alignManyLA2004} and \cite{alignEqualSpacing2010} worked with multi-frame slices.
As LGE data is single-frame, the available information that can be used for its realignment is $N_\mathrm{f}-1$ times fewer, where $N_\mathrm{f}$ is the number of available frames.

\subsection{Myocardium Segmentation}

The delineation of myocardial contours is a prerequisite to subsequent automatic localization and quantification of infarcts -- nearly all such works~\cite{infarctSeg2005,A_Comprehensive_Approach,infarctSeg2010a,infarctSeg2010b,infarctSeg2011} in the literature assume that high quality myocardium segmentation is given, either manually or (semi-) automatically.
However, automatic segmentation of the myocardium in LGE images is a difficult task.
There has been little research aimed at automatic myocardium segmentation for LGE CMR images, and there is no commercially or publicly available automatic segmentation tool for clinical use.
Most existing approaches utilize pre-delineated myocardial contours in cine CMR data of the same patient as \emph{a priori} knowledge~\cite{relatedworks01_Ciofolo2008,relatedworks02_Dikici2004,myownMICCAI2011}.
Such an approach is reasonable because the patient is asked to stay still during the entire acquisition process and there are many methods available for (semi-)automatic segmentation of the cine data, e.g., \cite{cineSegmentation_chen2008}.
The major difficulties of this approach include:
(i) displacement and nonrigid deformation between the cine and LGE data due to respiratory and body motion and imperfectness of ECG gating; and
(ii) different intensity characteristics of the cine and LGE data.

\subsection{Infarct Classification}

Early works on infarct classification were purely intensity-based with a global threshold, e.g., \cite{infarctSeg2005,FWHM2004}.
These methods suffer from the lack of spatial continuity, resulting in separated false positives probably caused by noise or artifacts.
To overcome this problem, \citet{FACT_I2006} proposed the feature analysis and combined thresholding (FACT) method, which enforces spatial constraints by assuming that the infarcts are generally sub-endocardial and reasonably large.
Recent works applied more advanced techniques to this problem.
\citet{heiberg2005semi} proposed a level set algorithm to regularize the thresholding and exclude small regions that constitute noise rather than infarction.
Similarly, \citet{metwally2010improved} proposed to enhance the thresholding using k-means clustering.
In \cite{A_Comprehensive_Approach} and \cite{infarctSeg2010a}, advanced signal intensity analysis is carried out based on the Rician distribution of noisy MRI data \cite{ricianMRI1995}, where the myocardium is modeled with a mixture model comprising a Rayleigh distribution for normal myocardium plus a Gaussian for infarcts.
However, these methods are generally 2D and hence do not take advantage of 3D spatial and intensity continuity.

Recently, \citet{infarctSeg2010b} proposed a method which first globally thresholds the image followed by false-positive and false-negative removal combining intensity and spatial information.
The method has three novelties:
(i) The LV BP is included as the same class as the infarcts during classification.
(ii) All SA slices of a patient are used together in the intensity distribution analysis to derive the global threshold.
This strategy works effectively when there is no obvious intensity inconsistency in a set of LGE images but \emph{degenerates} in the presence of significant intensity inconsistency.
(iii) The Otsu threshold \cite{OtsuThreshold1979} which maximizes the ratio of between-class and within-class variances is adopted, so there is no need to make any assumption regarding the underlying true distribution.
Nevertheless, this work suffers from two drawbacks: there is neither correction for intensity inconsistencies nor correction for misalignment artifacts across slices, and it is still based on a global threshold.

The only work in the literature that handled both misalignment artifacts and intensity inconsistencies was presented in \cite{infarctSeg2011}.
SA slices are realigned by aligning myocardial centers of all slices in a line;
however, this is not always correct since anatomically the slices' myocardial centers do not necessarily form a line.
The intensity of each slice is adjusted to the same range as in the first slice.
Although doing so can make the overall intensity characteristics of the entire images more consistent, it does not necessarily normalize the intensities of the LV, which despite being the real target only occupies a relatively small area in each image.
Another drawback of this work is that again global thresholding is performed as the major classification process.

\subsection{Comprehensive Frameworks}

To the best of our knowledge, the only existing comprehensive framework for automatic quantification of LGE CMR images is \cite{A_Comprehensive_Approach}, which conducted a joint analysis of the LGE and perfusion CMR images.
However, its segmentation of the myocardium in LGE images was done semi-interactively.
Although a solution for the misalignment artifacts of LGE slices was proposed, the navigator-gated whole-heart coronary angiography volume data, which was used as the reference for the realignment of LGE slices, is not widely available in usual clinical settings.
Finally, this work did not consider intensity inconsistencies across LGE slices and it is unclear whether the detection of infarcts was done in 2D or 3D.

\section{Method}
\label{sec:method}
The proposed framework comprises four steps:
(i) correction of misalignment artifacts;
(ii) 3D segmentation of the myocardium;
(iii) intensity normalization of the LV regions in a stack of SA slices; and
(iv) 3D classification of the myocardium into infarcted and normal regions.

\subsection{Correction of Misalignment Artifacts}

We propose an effective and robust method for misalignment correction of LGE CMR slices.
Unlike previous methods which merely rely on similarity measurement at intersections of slices, we propose to also utilize the innate physical continuity of the heart throughout the SA stack to establish interlinks within the SA slices and group them together.
This is achieved by introducing a novel cost term, namely, the contiguous cost, between adjacent SA slices.
The contiguous cost not only prevents unrealistic movement of the slices, but also improves accuracy of the misalignment correction.
%In addition, it will not be easily violated in practice because it is based on a physical property rather than an assumption.

In our method, we incorporate only translational but no rotational correction based on two research findings:
(i) a study on motion and deformation of the heart due to respiration found only very small rotations (typically a couple of degrees) \cite{heartMotionStudy}.
(ii) in another study \cite{alignEqualSpacing2010}, a method capable of full 3D translational and rotational corrections found no significant improvement on final results with the rotational correction.

\subsubsection{Intersecting Cost}
The intersecting cost refers to the dissimilarity measure along the intersecting line between a pair of intersecting slices.
Assume there are $m$ SA and $n$ LA slices (denoted by $I^{k}_{\mathrm{SA}}$ and $I^{j}_{\mathrm{LA}}$, where $k\in[1,m]$ and $j\in[1,n]$).
The position and orientation of a slice in the patient coordinate system are uniquely defined by two fields of the standard DICOM header -- ImagePositionPatient (IPP) and ImageOrientationPatient (IOP), respectively.
Given IPP and IOP of two slices, their intersection line can be computed.
Then, two line segments are sampled along the intersection line on both slices (Fig.~\ref{fig:misalignment_n_correction_n_intersection}, third column).
The smaller PixelSpacing (PS, from the DICOM header) of the two slices' is used as the step size during sampling with linear interpolation.
Finally, the intersecting cost between the two slices is defined as the dissimilarity measure of the two sampled line segments:
\begin{equation}\label{eq:intersectingErr}
    E_{\mathrm{int}}=\begin{cases}
        &E_{\mathrm{int}}(I^{k}_{\mathrm{SA}}\bigcap I^{j}_{\mathrm{LA}}) = \mathcal{C}_{\boldsymbol{s}}(\boldsymbol{s}^{k}_{\mathrm{SA}}, \boldsymbol{s}^{j}_{\mathrm{LA}}),\\
        &E_{\mathrm{int}}(I^{j}_{\mathrm{LA}}\bigcap I^{j'}_{\mathrm{LA}}) = \mathcal{C}_{\boldsymbol{s}}(\boldsymbol{s}^{j}_{\mathrm{LA}}, \boldsymbol{s}^{j'}_{\mathrm{LA}}),
    \end{cases}
\end{equation}
where $I^{k}_{\mathrm{SA}}\bigcap I^{j}_{\mathrm{LA}}$ denotes the intersection between the $k^\mathrm{th}$ SA and $j^\mathrm{th}$ LA slices, and $I^{j}_{\mathrm{LA}}\bigcap I^{j'}_{\mathrm{LA}}$ the intersection between the $j^\mathrm{th}$ and ${j'}^\mathrm{th}$ LA slices ($j\neq j'$), $\boldsymbol{s}^{k}_{\mathrm{SA}}$ and $\boldsymbol{s}^{j}_{\mathrm{LA}}$ the sampled line segments on SA and LA slices respectively, and $\mathcal{C}_{\boldsymbol{s}}$ a function measuring dissimilarity between two line segments.

\begin{figure*}
  \centering
  \includegraphics[width=.9\textwidth]{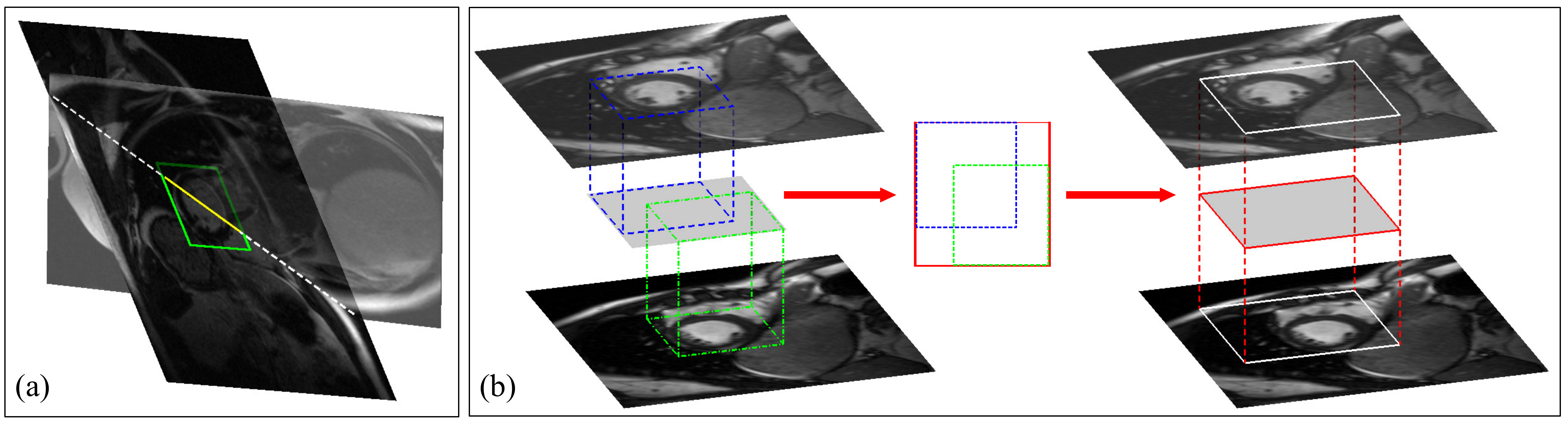}\\
  \caption{Illustration of the sampling ranges for
  (a) the intersecting cost: $\boldsymbol{s}^{k}_{\mathrm{SA}}$ and $\boldsymbol{s}^{j}_{\mathrm{LA}}$ are sampled along the intersection (dashed white) line only from the portion (yellow segment) lying within the SA slice's ROI (green rectangle);
  (b) the contiguous cost: the regions (white rectangles) in which $R^{k}_{\mathrm{SA}}$ and $R^{k+1}_{\mathrm{SA}}$ are sampled are defined by finding the smallest cuboid whose top and bottom faces are in the involved SA slices (planes) and can completely contain the corresponding ROI (dashed blue and green) in each slice.
  Distance between contiguous SA slices is exaggerated here for better visualization.}\label{fig:ROI}
\end{figure*}

$\boldsymbol{s}^{k}_{\mathrm{SA}}$ and $\boldsymbol{s}^{j}_{\mathrm{LA}}$ depend on IPP, IOP and PS of the two intersecting slices.
Of the three, IOP and PS can be considered constants as neither is changed by our method.
Consequently, $E_{\mathrm{int}}$ is actually a function of IPP.
Particularly, IPP of a slice stores the $x$, $y$ and $z$ coordinates of the slice in the patient coordinate system.
Thus, translation of a slice can be achieved conveniently by changing its IPP.
Because we are only interested in the LV, only a fraction of the intersection line between SA and LA slices is sampled to form $\boldsymbol{s}^{k}_{\mathrm{SA}}$ and $\boldsymbol{s}^{j}_{\mathrm{LA}}$.
The sampling range is confined within a region of interest (ROI) on each SA slice, which is predefined by identifying a bounding box that well contains the LV (Fig.~\ref{fig:ROI}(a)).
However, for the intersection line between different LA slices, the entire length is sampled to form $\boldsymbol{s}^{j}_{\mathrm{LA}}$ and $\boldsymbol{s}^{j'}_{\mathrm{LA}}$.
This is because if only the portion within the LV is sampled, there would be too little discriminative information (mostly from the BP) to determine relative positions of the LA slices.

\subsubsection{Contiguous Cost}
The contiguous cost refers to the discontinuity measure between two adjacent SA slices.
The LV is naturally smooth, and hence its representation throughout the stack of SA slices should be continuous\footnote{
Although there could sometimes be a gap between adjacent SA slices in clinical CMR data, this gap is always small
enough to obtain a correct 3D representation of the LV, and thus the smoothness of the LV is retained.}.
Although the continuity of the LV throughout the SA stack is an important physical property, it has been ignored by previous works.
We define the discontinuity measure between contiguous SA slices as the contiguous cost (as opposed to the intersecting  cost).
The contiguous cost is combined with the already defined intersecting cost to form the total cost to be minimized.
On one hand, the contiguous cost acts as a regularization to the intersecting cost, preventing it from pushing slices to unrealistic locations;
on the other hand, it can also improve the correction accuracy since it is derived from a strong physical property of the LV anatomy.

The discontinuity is measured by the dissimilarity between two regions of the same area sampled from adjacent SA slices.
Although the LV appears different in adjacent slices due to different imaging locations, the change is gradual.
Therefore, by minimizing the dissimilarity via translation of slices, the original relative positions of the SA slices can be recovered.
Again the ROI is predefined for each SA slice.
However, since the ROI is defined as the bounding box which well contains the LV, its size is varied for most SA slices as the LV size varies in most SA slices;
furthermore, during the translation, SA slices are deviated from each other, and so are their ROIs.
To solve this problem, we project the two ROIs of adjacent SA slices onto a middle plane and find the smallest rectangle which can completely contain the two projections on that plane, and then project this rectangle back onto the SA slices under consideration to determine the regions used for the discontinuity measure (Fig.~\ref{fig:ROI}(b)).
Noting that the projections are all along the normal to the SA planes, these procedures are actually finding the smallest cuboid whose top and bottom faces are in the SA slices (planes) under consideration and can completely contain the corresponding ROI in each slice.
Denoting the two regions sampled from the two adjacent SA slices by $R^{k}_{\mathrm{SA}}$ and $R^{k+1}_{\mathrm{SA}}$, the contiguous cost is defined as:
\begin{equation}\label{eq:contiguousCost}
    E_{\mathrm{cnt}}(I^{k}_{\mathrm{SA}}\parallel I^{k+1}_{\mathrm{SA}})=\mathcal{C}_{R}(R^{k}_{\mathrm{SA}}, R^{k+1}_{\mathrm{SA}}),\, k\in[1,m-1],
\end{equation}
where $I^{k}_{\mathrm{SA}}\parallel I^{k+1}_{\mathrm{SA}}$ denotes the adjacency of two consecutive SA slices, and $\mathcal{C}_{R}$ is a function measuring dissimilarity between two regions.
$E_{\mathrm{cnt}}$ is also a function of IPP, as the regions sampled on adjacent SA slices are determined by their IPPs (plus the constants IOP and PS).
Furthermore, because the translation of slices takes place in the patient coordinate system (instead of any of those coinciding with the SA plane), $E_{\mathrm{cnt}}$ is fully dependent on slices' translations in all $x$-, $y$-, and $z$- directions.

\subsubsection{Total Cost}
The total cost is a summation of all the intersecting and contiguous costs.
The intersecting cost is computed for all intersecting slices, including intersections between SA and LA slices, and between different LA slices;
each intersection is counted only once.
The contiguous cost is computed for all pairs of adjacent SA slices;
each adjacency is counted once too.
Now we can write the total cost as:
\begin{align}\label{eq:finalCost}
    &E_{\mathrm{align}}(\textit{IPP}_{\mathrm{all}}) =
    \sum^{m}_{k=1}\sum^{n}_{j=1} E_{\mathrm{int}}(I^{k}_{\mathrm{SA}}\bigcap I^{j}_{\mathrm{LA}})\ + \\
    &\sum^{n-1}_{j=1}\sum^{n}_{j'=j+1} E_{\mathrm{int}}(I^{j}_{\mathrm{LA}}\bigcap I^{j'}_{\mathrm{LA}})
    +\gamma\sum^{m-1}_{k=1} E_{\mathrm{cnt}}(I^{k}_{\mathrm{SA}}\parallel I^{k+1}_{\mathrm{SA}}), \nonumber
\end{align}
where $\textit{IPP}_{\mathrm{all}}=\{\textit{IPP}^{k}_{\mathrm{SA}},\ \textit{IPP}^{j}_{\mathrm{LA}} \mid k\in[1,m], j\in[1,n]\}$ represents $x$, $y$, $z$ coordinates of all slices under the patient coordinate system, and $\gamma$ a weighting factor.
$E_{\mathrm{align}}$ is minimized to find the optimal $\textit{IPP}^{\mathrm{opt}}_{\mathrm{all}}$ set, which is the corrected positions of all the slices.

For both $\mathcal{C}_{\boldsymbol{s}}$  and $\mathcal{C}_{R}$ we use the mean of the squared differences.
Noting that the intensities of the same structures may vary in different slices, the sampled $\boldsymbol{s}^{k}_{\mathrm{SA}}$, $\boldsymbol{s}^{j}_{\mathrm{LA}}$ and $R^{k}_{\mathrm{SA}}$ are pre-normalized to zero mean and unit standard deviation before being fed to $\mathcal{C}_{\boldsymbol{s}}$  and $\mathcal{C}_{R}$.
%The multiplier for ROI definition is 1.2, thus even when the bounding box is underestimated we can still have ROI covering the whole LV.
The ROI can be automatically estimated by a variant of the Hough transform tailored to the detection of annular shapes \cite{relatedworks01_Ciofolo2008} and our method is insensitive to small variations of the ROI.
The weighting factor $\gamma$ is set to 0.01 empirically.
The second row of Fig.~\ref{fig:misalignment_n_correction_n_intersection} shows two examples of the misalignment correction results.
We can see that the artifacts are largely reduced or even eliminated.

\subsection{3D Segmentation of the Myocardium}
\label{sec:method:subsec:myoSeg}
For high-quality automatic segmentation of the myocardium, we proposed a sophisticated 3D method.
Given myocardial contours in cine images as \emph{a priori} knowledge, either (semi-)automatically or manually obtained, the method initially propagates the \emph{a priori} segmentation from cine to LGE images via 2D translational registration.
Two \emph{simplex} meshes \cite{simplexmeshIJCV1999} representing respectively endocardial and epicardial surfaces are then constructed with the propagated segmentation contours.
After construction, the two meshes are nonrigidly deformed towards the myocardial edge points detected in both SA and LA LGE images in a unified 3D coordinate system.
Based on intensity characteristics of the LV in LGE images, we propose a novel parametric model of the LV for consistent myocardial edge point detection regardless of pathological status of the myocardium (infarcted or healthy) and of the type of the LGE images (SA or LA).
The final meshes after the nonrigid deformation are themselves a 3D segmentation of the myocardium.

Our method is distinct from the existing ones \cite{relatedworks01_Ciofolo2008,relatedworks02_Dikici2004,myownMICCAI2011} in two aspects.
(i) We integrate LA images (standard 4-chamber (4C) and 2-chamber (2C) views) with SA images for the 3D segmentation.
Besides providing complementary information about the LV between the largely spaced SA images, the LA images are also used for the correction of misalignment artifacts among slices.
(ii) We propose a novel parametric model of the LV for LGE images based on 1D intensity profiles.
This model is flexible to be applied to both SA and LA images, and self-adaptive to detect edge points of both infarcted and healthy myocardium.

We have evaluated the proposed method with the real patient data which is also used in this work.
The mean distance error between the automatic contours by our method and manual contours by the expert is $0.97\pm0.69$~mm, and the mean Dice coefficient (DC) \cite{Dice1945} for the myocardium is $87.60\pm8.63\%$.
The results confirm the observation that the method can generate accurate and reliable results for myocardial segmentation of LGE images.
We have also tested the robustness of the method with respect to varied \textit{a priori} segmentation.
In one of the experiments, we use manual and automatic cine segmentations as \textit{a priori} knowledge for a set of LGE data, and find that the segmentation results of the LGE images are similarity good: $1.02\pm0.18$ mm versus $1.12\pm0.29$~mm and $87.84\pm2.73\%$ versus $86.18\pm3.76\%$.
These results demonstrate that the method can compensate variations in the given \textit{a priori} knowledge and consistently produce accurate segmentations.

Due to the page limit, we will present the detailed description and comprehensive evaluation of this method in another paper.

\subsection{Correction of Intensity Inconsistencies}

\begin{figure*}[!t]
  \centering
  \includegraphics[width=\textwidth]{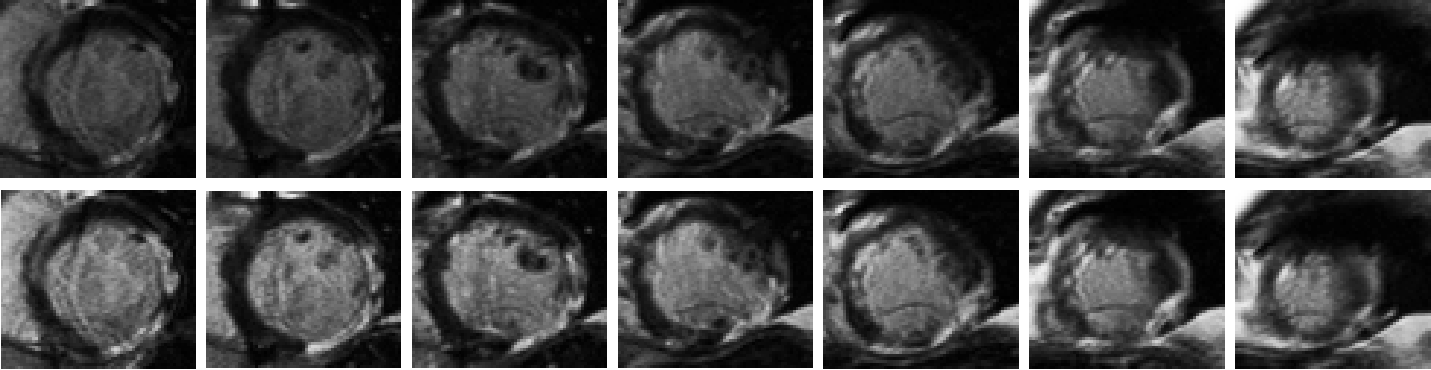}\\
  \caption{Illustration of the intensity normalization.
  First row: a stack of original SA images in one dataset; images become brighter from the mitral valve to the apex.
  Second row: the same image stack after the correction of intensity inconsistencies; the image intensities, especially in the LV regions, are more consistent.}\label{fig:intensityCorrectionComparison}
\end{figure*}

Signal intensities may vary from one LGE image to another even in the same dataset due to different time delays after injection of the contrast agent.
A usual pattern is that the intensities tend to become brighter from the mitral valve to the apex (see the first row of Fig.~\ref{fig:intensityCorrectionComparison}).
%For example, in some datasets normal myocardium in very apical slices may appear as bright as or close to infarcts in more basal slices.
Therefore, for any infarct classification method that combines multiple slices, normalization of the intensity across slices is necessary.
We propose a novel method which only considers local regional intensities of the LV and thus can effectively eliminate or reduce the intensity inconsistencies in the LV region.
The rationale underlying our normalization is that the intensity of the BP should be roughly the same across slices.
The BP region in each image can be simply obtained by masking the region enclosed by the endocardial contour.
However, dark papillary muscles are also considered as BP pixels in this way.
In order to normalize the BP and hence LV intensities more accurately, we have implemented an iterative algorithm based on the intensity distribution of the LV.

\subsubsection{Rician Distribution of the LV in LGE CMR Images}
In the presence of noise, the intensity of MR images is shown to be governed by a Rician distribution \cite{ricianMRI1995}.
In the case of the LV in LGE images, the Rayleigh distribution models normal myocardium (dark regions) while the Gaussian distribution models infarcts plus the BP (bright regions), and the entire LV is modeled with the mixture model:
\begin{equation}\label{eq:RicianDistribution}
\begin{split}
    &Rcn(x) = Pr_\mathrm{Ray}(x) + Pr_\mathrm{Gau}(x)\\
    &=\underbrace{\alpha_{\mathrm{R}}\frac{x}{\sigma_{\mathrm{R}}^{2}}\exp(-\frac{x^{2}}{2\sigma_{\mathrm{R}}^{2}})}_{\text{\footnotesize{Rayleigh distribution}}} +\underbrace{\alpha_{\mathrm{G}}\frac{1}{\sqrt{2\pi}\sigma_{\mathrm{G}}}\exp[-\frac{1}{2}(\frac{x-\mu}{\sigma_{\mathrm{G}}})^2]}_{\text{\footnotesize{Gaussian distribution}}},
\end{split}
\end{equation}
where $Pr_\mathrm{Ray}(x)$ and $Pr_\mathrm{Gau}(x)$ denote the Rayleigh and Gaussian distributions respectively, and $\alpha_\mathrm{R}$, $\delta_\mathrm{R}$, $\alpha_\mathrm{G}$, $\delta_\mathrm{G}$, $\mu$ their parameters.
The combination of the BP and infarcts as one class is not only reasonable because there is no significant difference between their intensity distributions \cite{infarctSeg2005}, but also helpful to make the assumed bimodal distribution more prominent \cite{infarctSeg2010b}.
During the acquisition of LGE CMR images, technologists try to null signals from the normal myocardium as much as possible in order to give prominence to the hyper-enhanced regions.
Consequently, the intensities of the normal myocardium is often shifted to the dark side with an unknown offset.
It has been suggested that introducing an extra offset parameter $a$ to the Rayleigh distribution would lead to more accurate estimation of (\ref{eq:RicianDistribution}) \cite{infarctSeg2010a}.
We follow this approach and thus $Pr_\mathrm{Ray}(x)$ becomes:
\begin{equation}
    Pr_\mathrm{Ray}(x) = \alpha_{\mathrm{R}}\frac{(x+a)}{\sigma_{\mathrm{R}}^{2}}\exp\left[-\frac{(x+a)^{2}}{2\sigma_{\mathrm{R}}^{2}}\right].
\end{equation}

\begin{figure}
  \centering
  \includegraphics[width=.95\columnwidth]{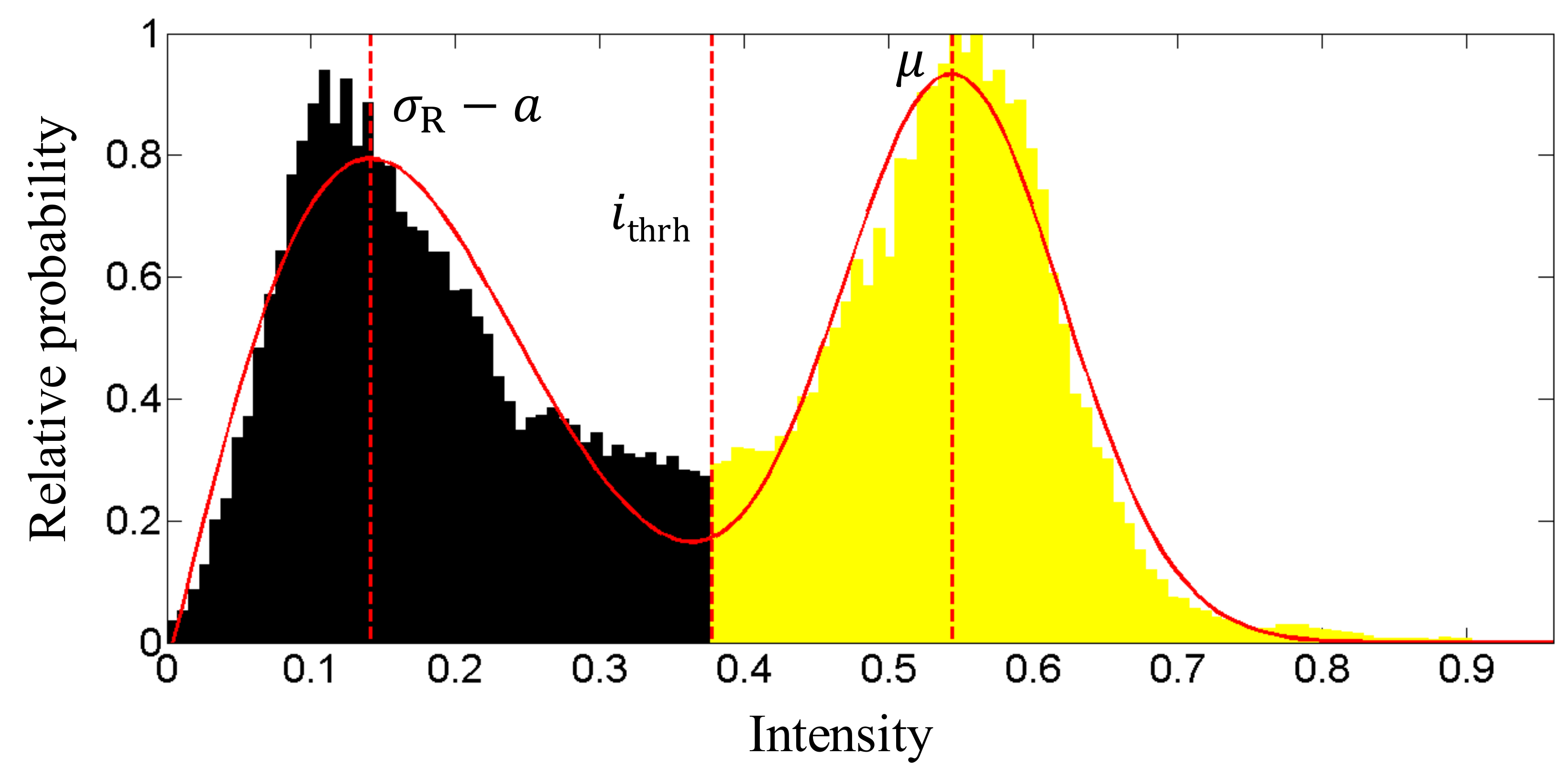}\\
  \caption{The fitted Rician distribution overlaid on the relative probability distribution.
  Also overlaid are vertical position lines of $\sigma_\mathrm{R}-a$, $i_\mathrm{thrh}$ and $\mu$.}\label{fig:RicianFit}
\end{figure}

We extract all the LV voxels from the regions enclosed by the epicardial contours in the stack of SA slices and obtain the intensity histogram.
We then obtain a \emph{relative probability} distribution by normalizing the histogram with the largest frequency count and fit the relative probability distribution to (\ref{eq:RicianDistribution}) to solve for $Rcn(x)$.
Figure~\ref{fig:RicianFit} shows an example of the fitted Rician distribution overlaid on the relative probability distribution.
The intersection of $Pr_\mathrm{Ray}(x)$ and $Pr_\mathrm{Gau}(x)$ is found and denoted as $i_\mathrm{thrh}$.
Thereafter BP voxels can be extracted from the regions enclosed by the endocardial contours but excluding those with intensities smaller than $i_\mathrm{thrh}$ (considered papillary muscles).
We take the middle slice in the SA stack as the reference and calculate the mean intensity of its BP pixels $i_\mathrm{m\_ref}$.
Then the rest of the slices are normalized by multiplying the factor $i_\mathrm{m\_ref} / i_\mathrm{m\_k}$, where $i_\mathrm{m\_k}$ is the mean intensity of the BP pixels in the $k^\mathrm{th}$ slice.
Finally, the whole stack of slices are normalized \emph{together} to the range $[0,1]$.

\subsubsection{Iterative Normalization}

Since the above described procedure may have changed the shape of the underlying intensity histogram of the stack of LGE slices, it is likely that if we repeat we would still find $i_\mathrm{m\_ref} / i_\mathrm{m\_k}$ unequal to 1 (as $i_\mathrm{thrh}$ is changed along with the histogram shape).
Therefore, we iterate the normalizing procedure until the histogram stabilizes, which is determined by examining how close the ratio $i_\mathrm{m\_ref} / i_\mathrm{m\_k}$ is to 1 for every slice.
% Although this iterative algorithm is simple in implementation but works effectively.
Figure~\ref{fig:intensityCorrectionComparison} shows slices of a dataset before and after the intensity normalization as a comparison.
The final $Pr_\mathrm{Ray}(x)$ and $Pr_\mathrm{Gau}(x)$ obtained when the iteration stops will be used further for infarct classification by the graph-cut algorithm.

\subsection{Classification of Infarcts by Graph-Cut}
\label{sec:subsec:graphCutClassify}

\subsubsection{3D Graph-Cut}
We employ the ND graph-cut algorithm presented in \cite{graphCutJolly2001} for our 3D classification of infarcts.
Let $p$, $q$ be any two different pixels in the image $I$ and $l_p$ the label assigned to $p$ in a two-label system $l_p\in\{0,1\}$, then a labeling set $L$ of every pixel in $I$: $L=\{l_p\ |\ p\in I\}$ represents a segmentation of $I$.
Based on the concept of maximum \textit{a posteriori} estimate in a locally dependent Markov random field \cite{MRFgraphcut1998}, the energy functional used by the graph-cut algorithm can be written as:
\begin{equation}
  E_\mathrm{seg}(L) = \lambda \sum_{p\in I} -\ln Pr(i_p\ |\ l_p) + \sum_{p\in I} \sum_{q\in\mathcal{N}_p} V_{(p,q)}(l_p,l_q),
\end{equation}
where $Pr(i_p\ |\ l_p)$ is the \emph{relative} likelihood, $V_{(p,q)}$ a \emph{clique potential}, $\mathcal{N}_p$ the set of all neighboring pixels of $p$, and $\lambda$ a positive weighting factor.
The term $-\ln Pr(i_p\ |\ l_p)$ denotes the cost of assigning label $l_p$ to the pixel $p$ of intensity $i_p$, and $V_{(p,q)}(l_p,l_q)$ denotes the cost of assigning labels $l_p$ and $l_q$, respectively, to the two neighboring pixels $p$ and $q$.

The input to the graph-cut algorithm is just the segmented myocardium, that is, regions other than the myocardium are masked out to eliminate irrelevant information.
Note that the input is a stack of 2D images, i.e., a 3D image.
Since there are only two allowable values for $l_p$, i.e., 1 for infarct and 0 for normal myocardium, $Pr(i_p\ |\ l_p)$ is defined as:
\begin{align}
    Pr(i_p\ |\ 1) &= \begin{cases}
        Pr_\mathrm{Gau}(i_p),\quad &\text{if}\quad i_p\leq \mu \\
        1,\quad                    &\text{if}\quad i_p>\mu
    \end{cases}\\
    Pr(i_p\ |\ 0) &= \begin{cases}
        Pr_\mathrm{Ray}(i_p),\quad &\text{if}\quad i_p\geq(\sigma_\mathrm{R}-a) \\
        1,\quad                    &\text{if}\quad i_p<(\sigma_\mathrm{R}-a)
    \end{cases}
\end{align}
The two special cases -- $Pr(i_p\ |\ 1)=1$ for $i_p$ larger than $\mu$ and $Pr(i_p\ |\ 0)=1$ for $i_p$ smaller than $(\sigma_\mathrm{R}-a)$ -- conforms to the clinical experience that a pixel brighter than the mode of the Gaussian distribution must be given the infarct label 1 and one darker than the mode of the Rayleigh distribution must be given the myocardium label 0.
In these cases, there should be no cost on the corresponding labelings, which is embodied by the fact that the logarithm of 1 is 0.

%We define an interaction potential analogous to what was used in \cite{graphCutJolly2001}:
%\begin{equation}\label{eq:graphCut_myVpq}
%  V_{(p,q)}(l_p,l_q) = |l_p-l_q| \cdot \exp\left(-\frac{(i_p-i_q)^2}{2\sigma^2}\right) \cdot \frac{1}{\mathrm{dist}(p,q)}.
%\end{equation}
%where $\mathrm{dist}(p,q)$ denotes the Euclidean distance between pixels $p$ and $q$.
%$|l_p-l_q|$ makes $V_{(p,q)}$ zero when $p$ and $q$ have the same label.
%Only when $l_p$ and $l_q$ are different will the interaction potential $V_{(p,q)}$ add a cost to $E_\mathrm{seg}(L)$.
%Therefore, $V_{(p,q)}$ acts only on those neighboring pixels with different labels, which are actually boundary pixels in the current labeling $L=\{l_p\}$.
%Moreover, when intensities of two pixels with different labels differ greatly, $V_{(p,q)}$ becomes small, indicating a potential edge of large intensity jump.

We also use the interaction potential $V_{(p,q)}$ proposed in \cite{graphCutJolly2001}.
However, instead of setting $\sigma$ -- the constant that controls the effective range of the potential well -- to be the estimated image noise, we set $\sigma=\mu-(\sigma_\mathrm{R}-a)$, which is the distance between the modes of $Pr_\mathrm{Ray}(x)$ and $Pr_\mathrm{Gau}(x)$.
$E_\mathrm{seg}(L)$ is minimized with the efficient $\alpha$-expansion algorithm \cite{alphaExpansion1999}.
Because only two labels are involved in our specific problem, the exact global minimum of $E_\mathrm{seg}(L)$ can be reached \cite{firstGraphCut}.
As for the definition of $\mathcal{N}_p$, we use the 6-connected neighborhood.
$\lambda$ is set to 1 at all times.
%Exemplary infarct classification result after the graph-cut is shown in Fig.~\ref{fig:classifySteps:afterGraphCut}.
%
%\begin{figure}[!t]
%    \centerline{
%    \hfil
%    \subfloat[]{
%        \includegraphics[width=.3\columnwidth]{classify_a}
%        }
%    \hfil
%    \subfloat[]{
%        \includegraphics[width=.3\columnwidth]{classify_b}
%        \label{fig:classifySteps:afterGraphCut}}
%    \hfil
%    \subfloat[]{
%        \includegraphics[width=.3\columnwidth]{classify_c}
%        \label{fig:classifySteps:final}}
%    \hfil
%    }
%    \caption{Illustration of the classification steps: (a) the LGE image with myocardial contours overlaid on; (b) classification result after the graph-cut; (c) final classification result after the post-processing.}\label{fig:classifySteps}
%\end{figure}

\subsubsection{Post-Processing}

False positive of infarcts is usually caused by inclusion of thin layers of epicardial fats or endocardial BP inside the myocardial contours.
This kind of contour tracing flaws are inevitable for both manual and \mbox{(semi-)} automatic delineations.
Sometimes patches of artifacts within the myocardium are another source of false positive.
Meanwhile, false negative may happen at regions with intermediate intensities, which are caused by the partial volume effect, i.e., the mixing of infarcted and normal myocardium.
Therefore, post-processing is necessary to remove these misclassifications.
Since there are already extensive works on the post-processing techniques, we simply follow those proposed in \cite{infarctSeg2010b} for the implementation, which have been proven effective and also work readily in our experiments.
To include MVO, sub-endocardial dark blobs surrounded by hyper-enhanced infarcts are located and classified as infarcts as well.

%An example of the final classification result after post-processing is shown in Fig.~\ref{fig:classifySteps:final}.

\subsection{Implementation}

Our framework is implemented under the \textsc{Matlab} R2011a environment, on a PC with Intel Core2 Duo T9400 processer, 3 GB RAM and running a 32-bit Microsoft Windows 7 OS.
Optimization of $E_{\mathrm{align}}$ and fitting of $Rcn(x)$ are achieved with the Optimization Toolbox and Curve Fitting Toolbox of \textsc{Matlab} R2011a, respectively.
The average running times for each step in the framework are: 26 s for the misalignment correction, 2.1 min for the 3D myocardium segmentation and 1.5 s for the intensity normalization plus 3D infarct classification.
The running times can be significantly reduced if the proposed framework is implemented in a more efficient programming language with multi-threading enabled.
The framework is found insensitive to the exact values of the parameters $\gamma$ and $\lambda$ by our preliminary experiments.

\section{Experimental Results and Discussion}
\label{sec:results}

\subsection{Data Specification and Experimental Settings}
\label{sec:subsec:dataSpecs}

20 sets of real patient data (denoted by DAT$_{1-20}$) with varying amounts of infarcts are used for validation of the proposed framework.
The data were acquired with ECG gating by a 1.5T Siemens Symphony MRI scanner from 20 patients (18 males and 2 females, 38-81 years old) three months after myocardial infarction, following a bolus injection of gadolinium-based contrast agent.
The sequence parameters used for the LGE image acquisition are as follows: flip angle 25$^\circ$, repetition time 650-1000 ms, echo time 4.18 ms, bandwidth 130 Hz/pixel, image width and height 176-256 pixels, in-plane isotropic pixel size 1.17-1.56 mm, slice thickness 7 mm and gap 3 mm.
Depending on the individual heart size, there could be 6-8 usable SA slices within the LV in each dataset;
standard 4C and 2C LA views were acquired along with the SA views.
In total 149 SA slices were included in the study.
The data were analyzed by medical experts (including segmentation of the myocardium and infarcts) manually and the results were used as the reference standard in our experiments.
The percentage of the infarcts with respect to the entire myocardium is commonly used by cardiologists to reflect the extent and severity of infarction.
According to the reference standard, volumetric infarct percentages for the 20 patients range from 0 to $32.31\%$ (two cases of non-infarct) with the mean and standard deviation of 18.60$\pm$10.43\%, with five patients having MVO.

For myocardium segmentation, the general quantitative evaluation results are already provided in Section \ref{sec:method:subsec:myoSeg}.
In this paper we focus on infarct classification and the final quantification results.
Apart from the infarct percentage (denoted by I/M\% afterwards), the DC is employed to measure the overlapping rates of infarcts classified automatically and manually.
I/M\% calculated from the manual and automatic results are compared by the Bland-Altman (BA) analysis \cite{Bland-Altman1986}.

\subsection{Quantification Results}

\begin{figure}
    \centering
    \includegraphics[width=.95\columnwidth]{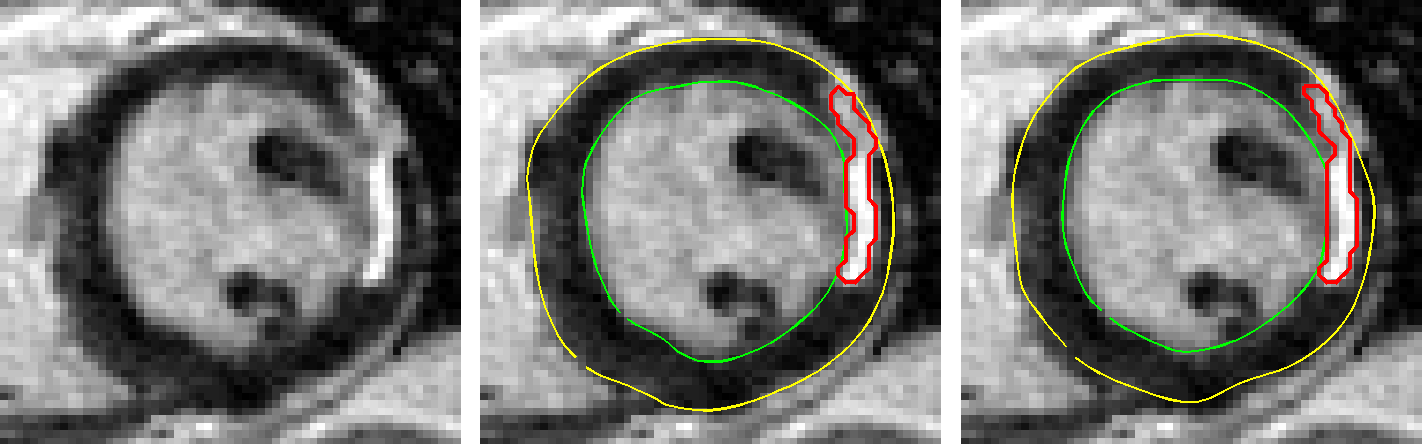}\\
    \includegraphics[width=.95\columnwidth]{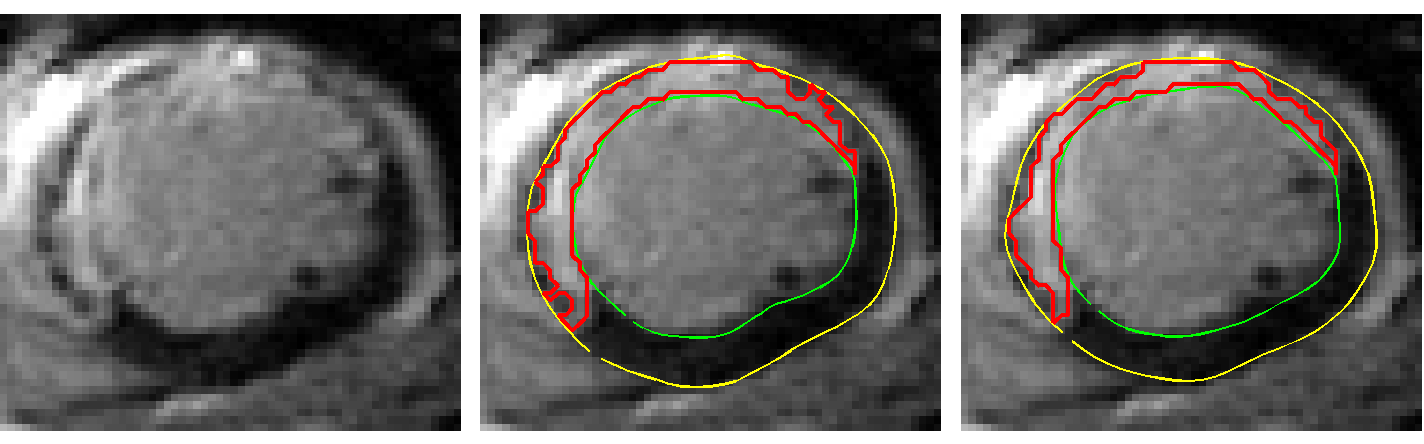}\\
    \includegraphics[width=.95\columnwidth]{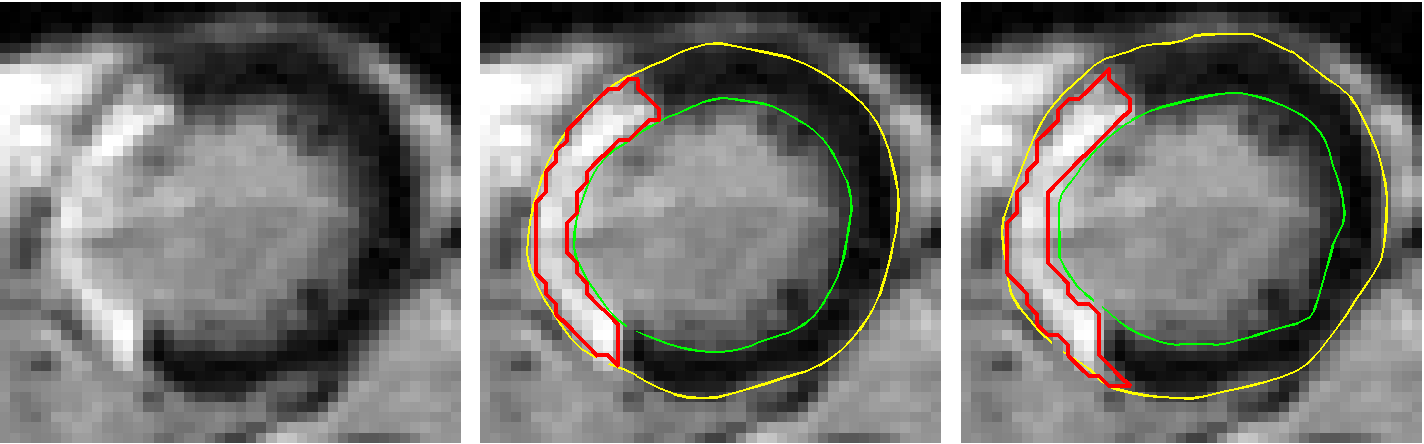}\\
    \begin{minipage}{.95\columnwidth}
        \centering\footnotesize
        \makebox[.32\linewidth][c]{(a)} \hfill \makebox[.32\linewidth][c]{(b)} \hfill \makebox[.32\linewidth][c]{(c)}
    \end{minipage}\\
    \caption{Exemplary results (each row shows a slice from a different subject): (a) the original image; (b) the proposed framework; (c) the reference standard.
    The green and yellow contours delineate the endocardium and epicardium respectively, and the red contours delineate the infarcts.}\label{fig:finalResults}
\end{figure}

Figure \ref{fig:finalResults} shows the automatic and manual results on 3 LGE images with different extents of infarction and different intensities.
Qualitatively we find that our framework consistently produces myocardial and infarct segmentations close to those by the experts, though several exceptions of failed cases are found in few very apical slices (e.g., the slice shown in Fig.~\ref{fig:failure}).

\subsubsection{Volumetric Analysis}
\begin{figure}
    \rotatebox{90}{\hspace{6mm} \footnotesize Automatic I/M\%$-$Manual I/M\%}
    \begin{minipage}[b]{.45\textwidth}
      \centering
      \includegraphics[width=\linewidth]{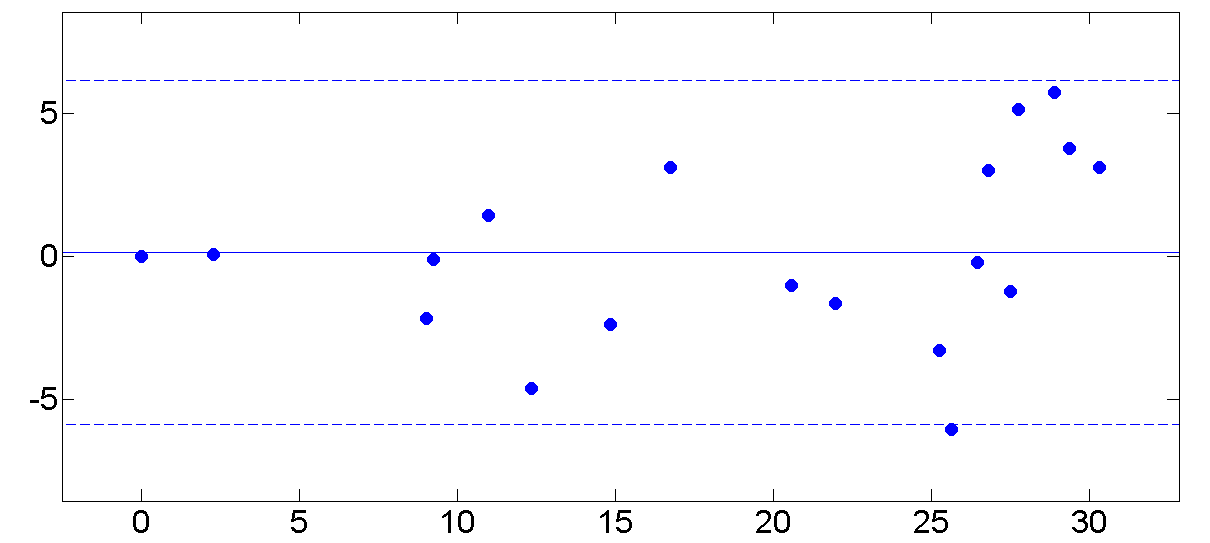}\\
      \footnotesize (Automatic I/M\%$+$Manual I/M\%)$/$2\vspace{3mm}
      \label{fig_first_case}
    \end{minipage}
    \\
    \rotatebox{90}{\hspace{2.5mm} \footnotesize Automatic I/M\%$-$Manual I/M\%}
    \begin{minipage}[b]{.45\textwidth}
      \centering
      \includegraphics[width=\linewidth]{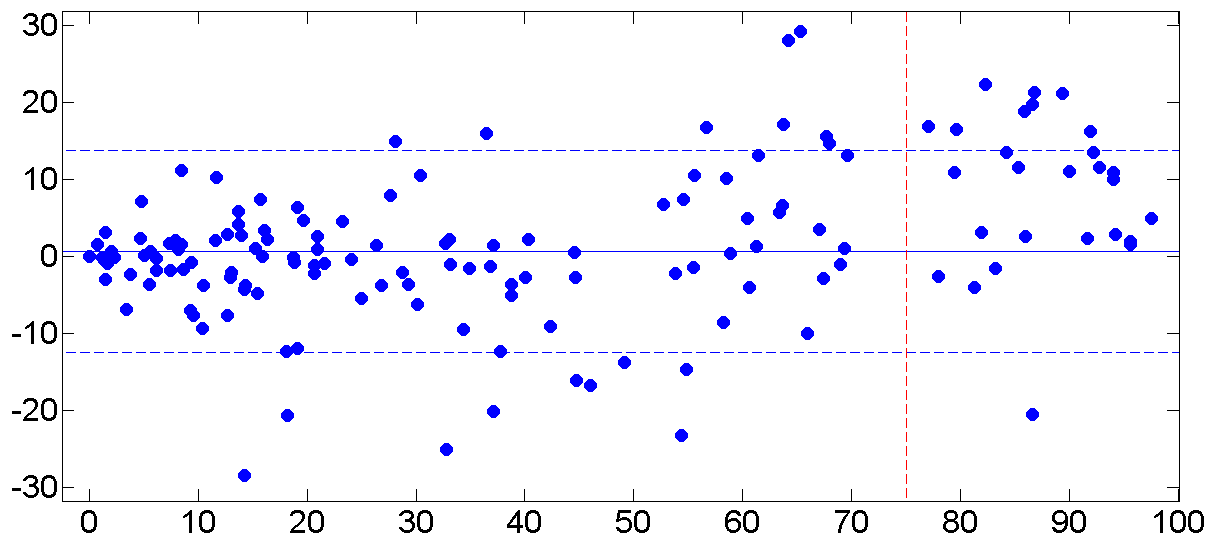}\\
      \footnotesize (Automatic I/M\%$+$Manual I/M\%)$/$2
      \label{fig_first_case}
    \end{minipage}

    \caption{Bland-Altman plots of volumetric (top) and AHA segment-wise (bottom) I/M\%.}\label{fig:Bland-Altman_plots}
\end{figure}

Of the 20 sets of data, 2 sets present no infarct after three months' recovery since the heart attack and our
framework correctly reported zero infarct.
The average DC for the other 18 sets of data is $80.90\pm5.04\%$, with the minimum and maximum values of $67.23\%$ and $88.25\%$, respectively.
Figure \ref{fig:Bland-Altman_plots} top shows the BA plot of the volumetric I/M\% (automatic minus manual results).
The BA analysis indicates that the automatic results report slightly higher percentages of infarcts than the manual results with a mean difference of $0.14\pm3.07\%$.
Furthermore, a Wilcoxon rank sum test indicates no significant difference between respective I/M\% ($P=0.95$).

\subsubsection{Segment-wise Analysis}
\begin{figure}
  \centering
  \rotatebox{90}{\hspace{5mm} \footnotesize{Infarct percentage}}
  \includegraphics[width=.9\columnwidth]{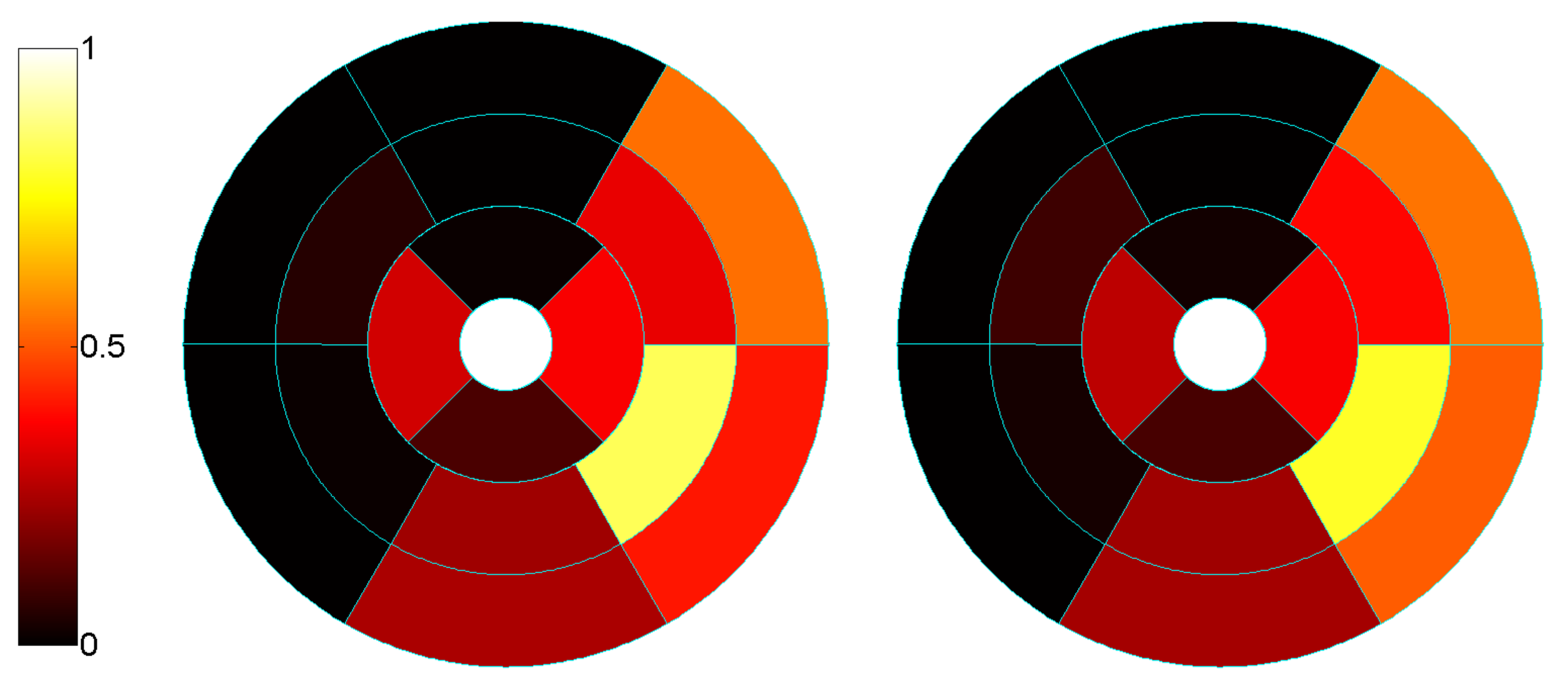}\\
  \leftline{\footnotesize
    \hspace{3.1cm}(a) \hspace{3.2cm}(b)
  }
  \caption{Bull's eye plots of I/M\%'s in standardized 16 segments for a set of LGE CMR data: (a) the automatic results; (b) the manual results. The extent of infarction is coded by the hot map. }\label{fig:BullseyePlot}
\end{figure}

The American Heart Association (AHA) has recommended to divide the LV myocardium into 17 segments for standardization \cite{AHA17segments};
since the 17$^\mathrm{th}$ segment, which is extracted from the apex in 4C and 2C LA views, is rarely used for infarct quantification by cardiologists in LGE CMR studies, only the 1$^\mathrm{st}$-16$^\mathrm{th}$ segments (denoted by Seg$_{1-16}$) extracted from SA views are included in our study.
As the segments are standardized with assignments to coronary arterial territories, examining segment-wise infarct presence can help identify the correlated vessel that is potentially occluded.
For easy visual interpretation, I/M\%'s of the 16 segments are plotted together in color-coded Bull's eye plots (Fig. \ref{fig:BullseyePlot}).

Compared to the LV volume-wise analysis, segment-wise analysis is more sensitive and expected to reveal more insights due to the employment of location-specific smaller volumes.
The BA analysis with all the $20 \times 16 = 320$ segments (Fig.~\ref{fig:Bland-Altman_plots} bottom) indicates that overall the automatic results report slightly higher I/M\% than the manual results, with a difference of $0.66\pm6.68\%$ ($P = 0.99$).
While the data points look compact around the zero level line at the smaller end of average I/M\%, the automatic results tend to report considerably more infarcts than the manual results when the average I/M\% is larger than 75\% (see the red dashed line in Fig.~\ref{fig:Bland-Altman_plots} bottom).

Of the 320 segments, 142 segments are deemed infarcted and 173 segments deemed free of infarct by both results.
Two false positives (i.e., the manual results reported zero while the automatic results reported positive I/M\%) are found with very small values: $3.06\%$ and $1.56\%$, respectively.
Meanwhile, three false negatives (i.e., the manual results reported positive while the automatic results reported zero I/M\%) are found and they are: $-3.03\%$, $-28.49\%$ and $-6.82\%$.
We have also compared absolute differences between I/M\% calculated from the automatic and manual results at different slice levels: basal (Seg$_{1-6}$), mid (Seg$_{7-12}$) and apical (Seg$_{13-16}$).
The multiple comparison following a Kruskal-Wallis test indicates that while there is no significant difference between basal ($2.14\pm5.21\%$) and mid ($3.12\pm5.47\%$) levels, the absolute differences at the apical level ($5.12\pm6.94\%$) are significantly larger than both of the basal and mid levels ($P\ll0.05$).

\subsection{3D Versus 2D Quantifications}

For 2D segmentation of the myocardium, we employed the method proposed in \cite{myownMICCAI2011}.
On the 20 sets of data, the mean distance error is $1.09\pm0.80$ mm, and the DC for the myocardium is $85.84\pm9.86\%$.
For 2D classification of infarcts, we applied the method presented in Section \ref{sec:subsec:graphCutClassify} to one slice at a time without the volumetric correction of intensity inconsistencies.
Also, the post processing took place in 2D.
As the 2D quantification works slice-wisely, naturally we compare the quantification results at the slice level.
The comparison is shown in Table \ref{tab:2d_vs_3d}.
The quantification results of our 3D framework excel those of the 2D methods in three of the four compared aspects, and are equally good in the last.
\begin{table}[h]
  \centering
  \caption{3D versus 2D quantifications. Note: the DCs were calculated with slices deemed infarcted by all manual, 2D and 3D results~($N=102$).}\label{tab:2d_vs_3d}
  \begin{tabular}{c|c|c|c}
    \hline
     & 2D & 3D & p-value \\
    \hline
    DC (\%) & $73.09\pm16.17$ & $79.11\pm15.50$ & $\ll0.05$ \\
    Absolute diff. in I/M\% & $5.64\pm6.54$ & $3.56\pm5.04$ & 0.01 \\
    No. false positives & 13 & 3 & -- \\
    No. false negatives & 1 & 1 & -- \\
    \hline
  \end{tabular}
\end{table}

\subsection{Discussion}

\subsubsection{Accuracy and Applicability of the Framework}

Experimental results on the basis of LV volumes indicate that our framework is capable of producing a reliable evaluation of the overall extent and severity of infarction for a patient.
The volumetric BA analysis reports a nearly zero discrepancy ($0.14\%$) between the percentages of infarcts derived from the automatic and manual results with a very narrow limit of agreement ($SD = 3.07\%$).
The volumetric DCs also suggest that the automatic and manual results largely agree with each other (mean DC $= 80.90\pm5.04\%$).
However, we should be cautious when applying the framework to evaluate the extent of infarction in each AHA segment and identify potential supplying arteries of stenosis.
Special attention should be paid to quantification results of segments in the apical level, i.e., Seg$_{13-16}$, since the Kruskal-Wallis test with the slice levels as the factor indicates that discrepancies between the automatic and manual results are significantly larger at the apical level than at the basal and mid levels ($5.12\pm6.94\%$ versus $2.14\pm5.21\%$ and $3.12\pm5.47\%$).
Further, the only significant false negative ($-28.49\%$) out of the five cases of false errors is Seg$_{15}$ of DAT$_5$.
Therefore, the framework can still be applied to segment-wise quantification of infarcts if careful attention is paid to apical segments and proper correction schemes (where necessary) are provided.

\subsubsection{Implications}

\begin{figure}
  \centering
  \includegraphics[width=.95\columnwidth]{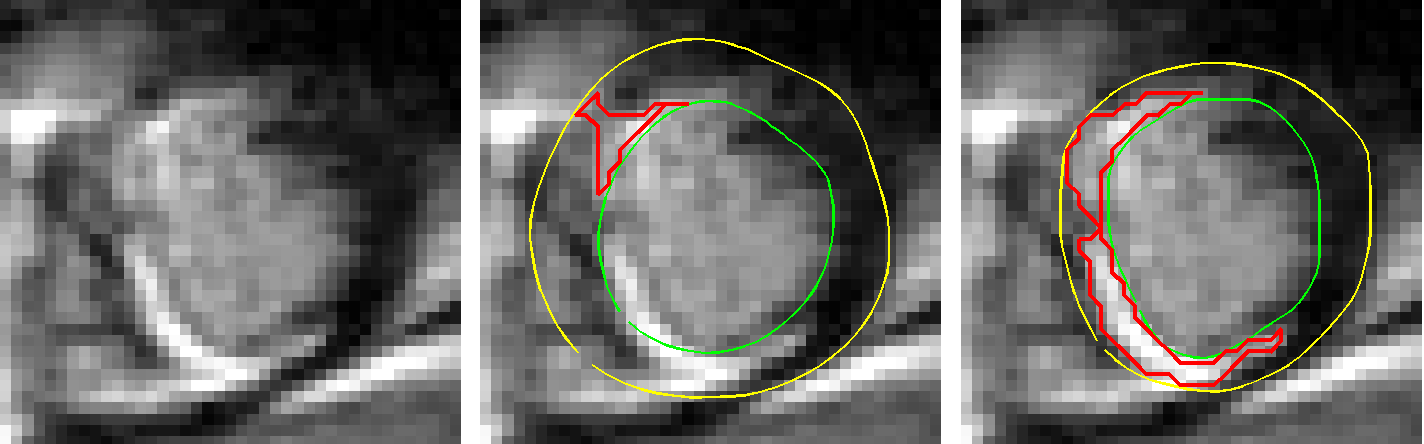}\\
  \begin{minipage}{.95\columnwidth}
        \centering\footnotesize
        \makebox[.32\linewidth][c]{(a)} \hfill \makebox[.32\linewidth][c]{(b)} \hfill \makebox[.32\linewidth][c]{(c)}
    \end{minipage}\\
  \caption{The most apical slice in DAT$_5$: (a) the original LGE image; (b) the automatic result; (c) the manual result.
  The incorrect classification of infarcts is caused by the failed automatic segmentation of the myocardium.}\label{fig:failure}
\end{figure}

As mentioned, we find that discrepancies between the automatic and manual results are significantly larger at the apical level than at the basal and mid levels.
This may be due to two reasons:
(i) Our framework tends to report considerably higher infarct percentages than the human observers when there are a large amount of infarcts.
Naturally there are significantly more infarcts at the apical than at the other two levels in real patients \cite{infarctSeg2005}, so our framework tends to report more infarcts for apical segments.
(ii) Automatic segmentation of the myocardium in apical slices is difficult, and myocardial contours of very poor quality cause failures anyway despite the good performance of the subsequent infarct classification method.
Such cases are illustrated in Fig.~\ref{fig:failure} with the most apical slice in DAT$_5$.
This failure occurs because our method utilizes myocardial contours in corresponding cine CMR images as the \textit{a priori} knowledge to guide the segmentation in LGE images.
When there is a large change in myocardium shape between corresponding cine and LGE images, the method may not be able to compensate and the segmentation fails.
Due to the use of ECG gating, such severe changes only happen at the extremely apical slices, where the LV is small and the apex's motion is large.
In fact the segmentation of the myocardium at apical slices is always so difficult that no method can consistently produce reliable segmentations.
Therefore, a user-correction scheme should be provided to handle such cases.
%It is also meaningful to basal and mid slices in the sense that experts can embody their experience in the final results with their modifications.

\subsubsection{Advantages of 3D Quantification}
The experimental results show that our 3D quantification framework outperform the 2D methods in the control group in three of the four examined aspects and has the same good performance in the last.
There are two reasons for this.
First, the myocardial contours produced by our 3D segmentation method is more accurate than those by the 2D method.
Second, due to the utilization of 3D spatial and intensity information in a stack of SA slices, the 3D classification of infarcts is also better than a 2D application of the same method to individual slices.
The latter has been confirmed by a subsequent comparative evaluation of merely the 3D and 2D \textit{classification} methods, in which the same manual myocardial contours drawn by the experts were input to both the 3D and 2D classification methods.
The results show that the 3D classification still performs better in the first three examined aspects in Table \ref{tab:2d_vs_3d} and equally well in the last.

\subsubsection{Limitations and Future Works}
One potential limitation of our framework is that the contiguous cost in the misalignment correction step does not work for sparse SA slices, e.g., when there is only one slice for each of the basal, mid and apical levels of the LV.
However, our misalignment correction method would still work for such cases by setting the weight for the contiguous cost $\gamma=0$, in which case the method is purely intersection based and thus quite similar to \cite{alignManyLA2004}.
Fortunately, most LGE CMR data nowadays have SA slices with slice spacings smaller than 1 cm (unless otherwise intentionally scanned in a sparse way) and our method utilizes this advantageous feature when available.
For future work, it would be interesting to study the impacts of different slice spacings and pixel sizes on the entire framework.

The evaluation in this work was conducted with limited (i.e., 20 sets) LGE CMR data.
If more datasets are involved, the framework could be tested by more potential variations of the human heart anatomy.
Moreover, reliable Kruskal-Wallis tests with the AHA segment order as the factor would be possible with a considerably larger database, other than the slice level used in this work.
This would provide insights specific to every AHA segment.
Another drawback of this work is that we did not explore the potential variations between different human observers.
As a study reported volumetric DCs of $80\pm8\%$ between the infarcts segmented by two observers given the same myocardial contours \cite{infarctSeg2010b}, we expect that, if the observers draw and use their respective delineations of the myocardium, the DC would drop further.
%Also, it would be interesting to compare the intra-observer variations with the accuracy of our framework.

It has recently been reported that the infarcts may not always be homogeneous and that tissue heterogeneity (core and peri-infarct zones) has great diagnostic and predictive potentials~\cite{yan2006characterization,schmidt2007infarct}.
In future work, we plan to extend our framework to also support the quantification of the infarct core and peri-infarct zones within the identified infarcts.
Finally, though the 17$^{th}$ segment in the AHA nomenclature, which stands for the apex in the LA views, is rarely used for the quantification of infarcts, it can provide valuable information on the presence and transmurality of the infarcts right at the apex.
Hence we are currently extending our 3D segmentation method to the apex in 4C and 2C LA views to support the analysis of infarcts there.
%Another interesting research direction is the joint analysis with other types of CMR images, e.g., tagged CMR.
%Combining the strains derived from tagged CMR and the infarcts identified from LGE CMR would provide more insights than analyzing either alone.

\section{Conclusion}
\label{sec:conclusion}

This paper presents a complete 3D framework for automatic quantification of LGE CMR images.
This framework achieves 3D segmentation of the myocardium as well as 3D classification of infarcts within the segmented myocardium, with robust and effective pre-processing measures overcoming misalignment artifacts and intensity inconsistencies across slices.
Qualitative and quantitative evaluations using real patient data demonstrated that the proposed framework is able to produce accurate segmentation and classification results and has the potential to be developed further as a clinical tool to generate objective quantification of LGE CMR images.

% An example of a floating table. Note that, for IEEE style tables, the
% \caption command should come BEFORE the table. Table text will default to
% \footnotesize as IEEE normally uses this smaller font for tables.
% The \label must come after \caption as always.
%
%\begin{table}[!t]
%% increase table row spacing, adjust to taste
%\renewcommand{\arraystretch}{1.3}
% if using array.sty, it might be a good idea to tweak the value of
% \extrarowheight as needed to properly center the text within the cells
%\caption{An Example of a Table}
%\label{table_example}
%\centering
%% Some packages, such as MDW tools, offer better commands for making tables
%% than the plain LaTeX2e tabular which is used here.
%\begin{tabular}{|c||c|}
%\hline
%One & Two\\
%\hline
%Three & Four\\
%\hline
%\end{tabular}
%\end{table}

% Note that IEEE does not put floats in the very first column - or typically
% anywhere on the first page for that matter. Also, in-text middle ("here")
% positioning is not used. Most IEEE journals use top floats exclusively.
% Note that, LaTeX2e, unlike IEEE journals, places footnotes above bottom
% floats. This can be corrected via the \fnbelowfloat command of the
% stfloats package.

% or


\appendices

%\section{}

\section*{Acknowledgment}

The authors would like to thank the Academic Research Fund, National University of Singapore, Ministry of Education, Singapore for funding the CMR studies. We are also grateful to the radiographers and staff at the Department of Diagnostic Imaging, National University Hospital, Singapore, for helping with the CMR scans.

\ifCLASSOPTIONcaptionsoff
  \newpage
\fi



\bibliographystyle{IEEEtranN}
\end{document}